\documentclass[a4paper,fleqn,usenatbib,letters]{mnras}

\pdfoutput=1

\usepackage{amssymb,amsmath,latexsym,mathrsfs}
\usepackage{graphicx,subfigure}
\usepackage{epsfig}
\usepackage{varioref,xr-hyper}
\usepackage{color}
\usepackage{multirow}
\usepackage{array}
\usepackage{hyperref}
\usepackage{wasysym}
\usepackage{color}
\usepackage{float}
\usepackage[utf8]{inputenc}
\usepackage[T1]{fontenc}

\title[Implications of the NANOGrav results for inflation]{Implications of the NANOGrav results for inflation}

\author[S. Vagnozzi]{
Sunny Vagnozzi\thanks{E-mail: \href{mailto:sunny.vagnozzi@ast.cam.ac.uk}{sunny.vagnozzi@ast.cam.ac.uk}}\thanks{Newton-Kavli Fellow}
\\
Kavli Institute for Cosmology, University of Cambridge, Madingley Road, Cambridge CB3 0HA, United Kingdom
}

\date{Accepted XXX. Received YYY; in original form ZZZ}

\pubyear{2020}

\begin{document}
\label{firstpage}
\pagerange{\pageref{firstpage}--\pageref{lastpage}}
\maketitle

\begin{abstract}
The NANOGrav pulsar timing array experiment reported evidence for a stochastic common-spectrum process affecting pulsar timing residuals in its 12.5-year dataset, which might be interpreted as the first detection of a stochastic gravitational wave background (SGWB). I examine whether the NANOGrav signal might be explained by an inflationary SGWB, focusing on the implications for the tensor spectral index $n_T$ and the tensor-to-scalar ratio $r$. Explaining NANOGrav while complying with upper limits on $r$ from \textit{BICEP2}/\textit{Keck Array} and \textit{Planck} requires $r \gtrsim {\cal O}(10^{-6})$ in conjunction with an extremely blue tensor spectrum, $0.7 \lesssim n_T \lesssim 1.3$. After discussing models which can realize such a blue spectrum, I show that this region of parameter space can be brought in agreement with Big Bang Nucleosynthesis constraints for a sufficiently low reheating scale, $T_{\rm rh} \lesssim 100\,{\rm GeV}-1\,{\rm TeV}$. With the important caveat of having assumed a power-law parametrization for the primordial tensor spectrum, an inflationary interpretation of the NANOGrav signal is therefore not excluded.
\end{abstract}


\section{Introduction}
\label{sec:introduction}

Several well-motivated physical scenarios predict the existence of a stochastic gravitational wave background (SGWB) spanning a wide range of frequencies~\citep[][]{Caprini:2018mtu,Christensen:2018iqi}, whose detection would provide an unique window onto physics operating at energy scales which cannot be reached on Earth. While some possible SGWB sources are more astrophysical in nature, \textit{e.g.} a cosmic population of inspiraling supermassive black hole binaries (SMBHBs)~\citep{Sesana:2004sp}, others are instead more cosmological in nature, \textit{e.g.} relics from cosmological phase transitions~\citep{Caprini:2010xv,Kobakhidze:2017mru,Ramberg:2019dgi}, such as cosmic strings~\citep[see e.g.][]{Siemens:2006yp,Blanco-Pillado:2017rnf}.

Gravitational waves (GWs) produced in the early Universe from the super-adiabatic amplification of zero-point quantum fluctuations of the gravitational field by inflation are a possible important SGWB source. Introduced to elegantly address the flatness, horizon, and monopole problems~\citep[][]{Kazanas:1980tx,Starobinsky:1980te,Sato:1981ds,Guth:1980zm,Mukhanov:1981xt,Linde:1981mu,Albrecht:1982wi}, inflation is a postulated stage of early quasi-de Sitter expansion which provides a compelling mechanism for the generation of a nearly Gaussian and nearly scale-invariant spectrum of scalar perturbations. This picture is remarkably consistent with Cosmic Microwave Background (CMB) and large-scale structure observations~\citep{Aghanim:2018eyx,Chowdhury:2019otk,Sabti:2020ser}, although difficulties with regards to embedding the simplest inflationary models within string-consistent UV completions has also been highlighted~\citep{Obied:2018sgi,Achucarro:2018vey,Garg:2018reu,Kehagias:2018uem,Kinney:2018nny,Geng:2019phi,Trivedi:2020wxf}.

The inflationary SGWB spans a wide frequency range~\citep{Maggiore:1999vm}, with the low-frequency range ($f \lesssim 10^{-15}\,{\rm Hz}$) best suited for detection via anisotropies in the CMB polarization field~\citep{Kamionkowski:2015yta}. Pulsar timing array (PTA) experiments, studying spatially correlated fluctuations induced by the SGWB on the arrival times of radio pulses from millisecond pulsars, and exploiting the fact that these objects behave as extremely stable clocks~\citep{1978SvA....22...36S,Detweiler:1979wn,1990ApJ...361..300F}, are sensitive to a SGWB in the nHz range.

The North American Nanohertz Observatory for Gravitational Waves (NANOGrav) has collected pulsar timing data since 2004, and recently found strong evidence for a stochastic common-spectrum process, strongly preferred over independent red-noise processes, from the analysis of their 12.5-year PTA dataset~\citep[][]{Arzoumanian:2020vkk}. If due to a SGWB, the detection is compatible with a GW signal with strain amplitude $A \sim 10^{-15}$ at $f \sim 3 \times 10^{-8}\,{\rm Hz}$, and a nearly flat spectrum. The detected signal is in tension with previous PTA SGWB upper limits in the same frequency range, a discrepancy attributed to an improved treatment of the intrinsic pulsar red-noise. However, no convincing detection of spatial quadrupolar correlations among the pulsar timing residuals, characterized by the so-called Hellings-Downs curve~\citep{Hellings:1983fr} and necessary to claim a SGWB detection, was achieved.

With these caveats in mind, it is nonetheless tempting to speculate as to the possible cosmological origin of the NANOGrav signal. Recent works have examined the possibility that this might be due to GWs generated by a network of cosmic strings~\citep{King:2020hyd,Ellis:2020ena,Blasi:2020mfx,Buchmuller:2020lbh,Samanta:2020cdk}, by phase transitions~\citep{Nakai:2020oit,Addazi:2020zcj,Ratzinger:2020koh,Bian:2020bps,Neronov:2020qrl,Li:2020cjj}, by dynamics of axion-like fields~\citep{Ratzinger:2020koh,Namba:2020kij}, or associated to the formation of primordial black holes~\citep{Vaskonen:2020lbd,DeLuca:2020agl,Kohri:2020qqd,Sugiyama:2020roc}.

In this \textit{Letter}, I address one of the simplest questions related to the NANOGrav result: ``\textit{Can the signal be due to an inflationary SGWB, and if so, what are the implications for inflationary parameters?}'' With a few caveats, I find that the answer to the first part of the question is ``\textit{Yes}''.

\section{Inflationary gravitational waves}
\label{sec:gwinflation}

I now review the description of GWs produced during inflation on the scales relevant for PTA experiments, closely following~\cite{Zhao:2013bba,Liu:2015psa}. I consider a spatially flat FLRW metric, whose perturbed line element in synchronous gauge reads:
\begin{eqnarray}
ds^2 = a^2(\eta) \left [ d\eta^2 - (\delta_{ij}+h_{ij})dx^idx^j \right ]\,,
\label{eq:flrw}
\end{eqnarray}
with $a$ and $\eta$ denoting scale factor and conformal time respectively, whereas the transverse and traceless part of the symmetric $3 \times 3$ matrix $h_{ij}$ describes GWs. Moving to Fourier space, focusing on one particular polarization and assuming isotropy, the GW field $h_k$ satisfies the following equation:
\begin{eqnarray}
h_k^{''}+2{\cal H}h_k^{'}+k^2h_k=0\,,
\label{eq:hdotdot}
\end{eqnarray}
with $^{'}$ denoting a derivative with respect to conformal time, and $k$ denoting the mode wavenumber. Consider now a GW field given at an initial conformal time $\eta_i$ by $h_k(\eta_i)$, characterized by its primordial spectrum $P_t(k) = 2k^3 \vert h_k(\eta_i) \vert^2/\pi^2$, and by its transfer function $T(\eta\,,k) = h_k(\eta)/h_k(\eta_i)$, with $h_k(\eta)$ computed by solving Eq.~(\ref{eq:hdotdot}) and evaluated at a conformal time $\eta>\eta_i$. The quantity relevant for GW detection experiments is the density parameter $\Omega_{\rm GW}$, \textit{i.e.} the ratio between the GW energy density today and the critical density:
\begin{eqnarray}
\Omega_{\rm GW} = \frac{1}{12H_0^2} \int dk \, \frac{P_t(k)\Dot{T}(\eta_0\,,k)}{k}\,,
\label{eq:omega}
\end{eqnarray}
where $\eta_0$ denotes conformal time today and the dot denotes a time derivative.For standard references explaining in more detail the relation between $h_{ij}$ in Eq.~(\ref{eq:flrw}), $h_k$ in Eq.~(\ref{eq:hdotdot}), and $\Omega_{\rm GW}$ in Eq.~(\ref{eq:omega}), the reader can refer to~\cite{Maggiore:1999vm,Boyle:2005se,Guzzetti:2016mkm}.

Connecting to inflation requires specifying the form of the primordial spectrum of tensor perturbations produced by inflation, $P_t(k)$, usually parametrized as a power-law:
\begin{eqnarray}
P_t(k) = rA_s(k_{\star}) \left ( \frac{k}{k_{\star}} \right )^{n_T}\,,
\label{eq:p}
\end{eqnarray}
where $r$ is the tensor-to-scalar ratio, $A_s$ is the amplitude of primordial scalar perturbations at the pivot scale $k=k_{\star}=0.05\,{\rm Mpc}^{-1}$, and $n_T$ is the tensor spectral index, with $n_T=0$ corresponding to a scale-invariant spectrum, and $n_T<0$ [$n_T>0$] corresponding to a red [blue] spectrum. Single-field slow-roll models satisfy the ``consistency relation'' $n_T=-r/8$~\citep{Copeland:1993ie}, leading to the expectation that the tensor spectrum should be slightly red-tilted.

Several works have shown that the transfer function $T(\eta_0\,,k)$ admits a relatively simple analytical approximation, assuming that inflation is followed by the standard radiation-, matter-, and dark energy-dominated eras~\citep{Turner:1993vb,Chongchitnan:2006pe,Zhao:2006mm,Giovannini:2009kg,Zhao:2013bba}. At PTA scales, the corresponding modes are much larger than the matter-equality wavenumber: $k \gg k_{\rm eq} \approx 0.073\Omega_mh^2\,{\rm Mpc}^{-1}$, with $\Omega_m$ and $h$ the matter density parameter and reduced Hubble constant. In this regime and adopting the aforementioned analytical approximation, the GW energy density today given by Eq.~(\ref{eq:omega}) reduces to~\citep{Zhao:2013bba}:
\begin{eqnarray}
\Omega_{\rm GW} \approx \frac{15}{16}\frac{\Omega_m^2rA_s}{H_0^2\eta_0^4k_{\rm eq}^2}\left ( \frac{k}{k_{\star}} \right )^{n_T}\,,
\label{eq:omega1}
\end{eqnarray}
which can be converted to the more useful frequency domain, $\Omega_{\rm GW}(f)$, through the relation $k=2\pi f$.

\section{The NANOGrav detection}
\label{sec:nanograv}

PTA GW search results are typically reported in terms of a pulsar timing-residual cross-power spectral density $S(f) \propto f^{-\gamma_{_{\rm CP}}}$. The results are also reported in terms of $h_c(f)$, the power spectrum of the characteristic GW strain, approximated as a power-law at the reference frequency $f_{\rm yr} = {\rm yr}^{-1}$:
\begin{eqnarray}
h_c(f) = A_{\rm CP} \left ( \frac{f}{f_{\rm yr}} \right )^{\alpha_{_{\rm CP}}}\,,
\label{eq:strainpowerspectrum}
\end{eqnarray}
where $\alpha_{_{\rm CP}}=(3-\gamma_{_{\rm CP}})/2$, with $\alpha_{_{\rm CP}}=-2/3$ [$\gamma_{_{\rm CP}}=13/3$] for the SMBHBs case. Moreover, $h_c(f)$ is related to $\Omega_{\rm GW}(f)$ given in Eq.~(\ref{eq:omega1}) through:
\begin{eqnarray}
\Omega_{\rm GW}(f) = \frac{2\pi^2}{3H_0^2}f^2h_c^2(f)\,.
\label{eq:hcomega}
\end{eqnarray}
PTA GW search results are reported via joint $A_{_{\rm CP}}$-$\gamma_{_{\rm CP}}$ posterior distributions, or the posterior distribution of $A_{_{\rm CP}}$ at a fiducial value of $\gamma_{_{\rm CP}}$ (typically $\gamma_{_{\rm CP}}=13/3$).

To connect to inflationary GWs, it is useful to relate $A_{_{\rm CP}}$ and $\gamma_{_{\rm CP}}$ to the inflationary parameters $n_T$ and $r$ by combining Eqs.~(\ref{eq:omega1}-\ref{eq:hcomega}). Inserting the best-fit values for the relevant cosmological parameters as determined by the \textit{Planck} collaboration~\citep{Aghanim:2018eyx} yields~\citep{Zhao:2013bba}:
\begin{eqnarray}
A_{_{\rm CP}} \approx 0.9\sqrt{r} \times 10^{5n_T-18}\,, \quad \gamma_{_{\rm CP}} = 5-n_T\,,
\label{eq:agammarnt}
\end{eqnarray}
with $A_{_{\rm CP}}$ at the reference frequency $f_{\rm yr}$ and $r$ at the pivot scale $k_{\star}$. From Eq.~(\ref{eq:agammarnt}) one sees that if the primordial tensor power spectrum is red as expected from single-field slow-roll inflationary models, the inflationary SGWB amplitude on PTA scales is undetectably small, with $A \lesssim 10^{-18}$.

\begin{figure}
\includegraphics[width=0.9\linewidth]{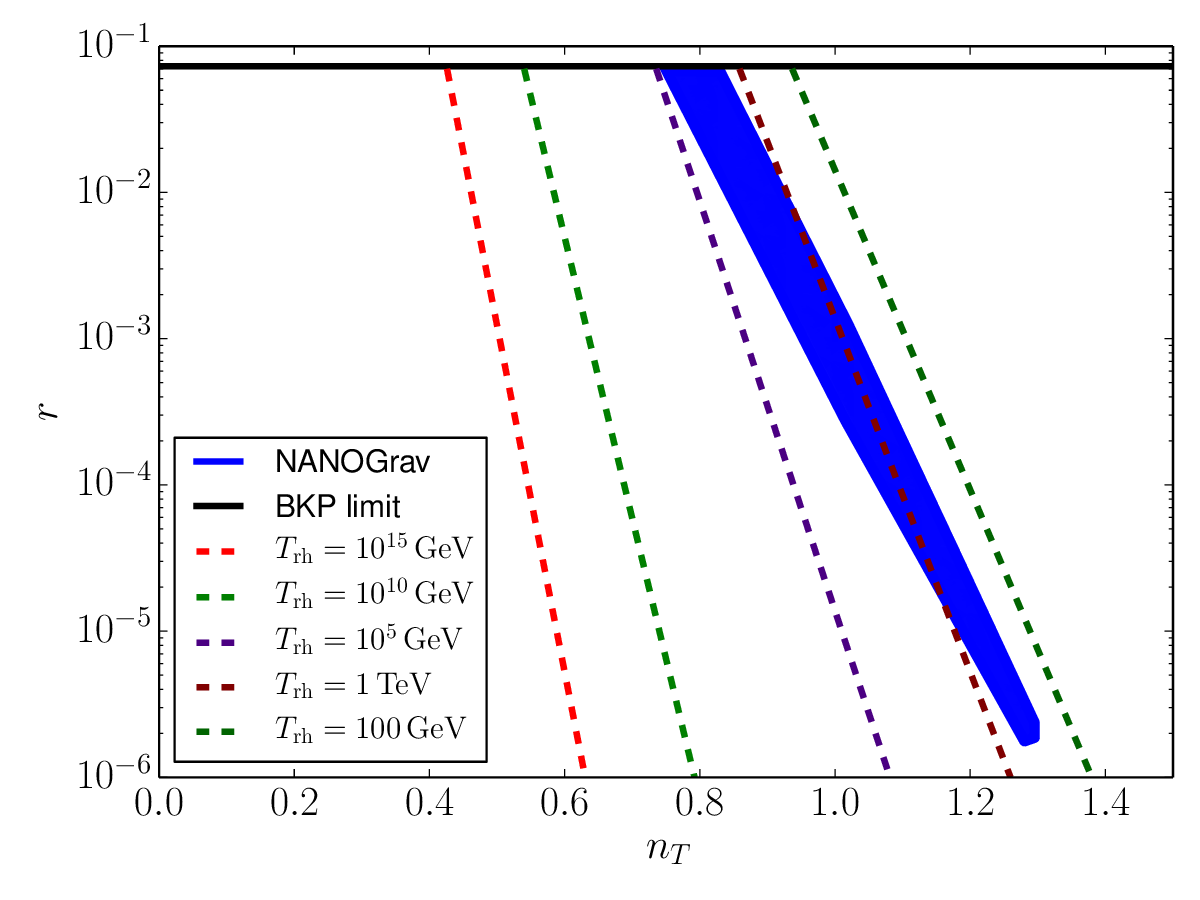}
\caption{Blue-shaded contour: region of $n_T$-$r$ parameter space required to explain the NANOGrav result assuming it is due to an inflationary SGWB. Solid horizontal line: limit $r<0.07$ set by \textit{BICEP2/Keck Array} and \textit{Planck}. Diagonal dashed lines: BBN limits for different values of the reheating temperature $T_{\rm rh}$ (regions to the right excluded, see legend for color coding).}
\label{fig:nanograv_fig1}
\end{figure}

The NANOGrav collaboration fitted the power-law approximation to the strain power spectrum to their 5 lowest frequency bins with highest signal-to-noise (see Fig.~13 in~\cite{Arzoumanian:2020vkk}), in the range $f \in (2.5 \times 10^{-9}; 9.0 \times 10^{-8})\,{\rm Hz}$, obtaining joint constraints on $A_{_{\rm CP}}$ and $\gamma_{_{\rm CP}}$ shown in the right panel of Fig.~1 of~\cite{Arzoumanian:2020vkk}. At 68\%~confidence level (C.L.) these indicate approximately $\log_{10}A_{_{\rm CP}} \in (-15.8 ; -15.0)$ and $\gamma_{_{\rm CP}} \in (4.5 ; 6.5)$, with the SMBHBs case $\gamma_{_{\rm CP}}=13/3 \approx 4.3$ falling just outside the 68\%~C.L. contours.

\section{Discussion and caveats}
\label{sec:implications}

I now identify the region of $n_T$-$r$ parameter space which can, at face value, explain the NANOGrav detection if interpreted as arising from an inflationary SGWB. I perform a scan of $n_T \in (-2 ; 2)$ and $\log_{10}r \in (-6 ; 0)$, and compute the resulting values of $A_{_{\rm CP}}$ and $\gamma_{_{\rm CP}}$ following Eq.~(\ref{eq:agammarnt}), retaining only points for which the corresponding $[\gamma_{_{\rm CP}}(n_T,r);A_{_{\rm CP}}(n_T,r)]$ pairs fall within the 95\%~C.L. contours corresponding to the NANOGrav detection. I impose the constraint $r<0.07$ at 95\%~C.L. from \textit{BICEP2}/\textit{Keck Array} and \textit{Planck} (\textit{BKP}) CMB data~\citep{Ade:2018gkx}.

The result is given by the blue-shaded region in Fig.~\ref{fig:nanograv_fig1}. The NANOGrav detection is compatible with a rather limited strip of $n_T$-$r$ parameter space, indicating approximately $n_T \in (0.7 ; 1.3)$ and $r \in (10^{-6} ; 0.07)$, with the upper limit on $r$ completely determined by the \textit{BKP} constraint. I find that the primordial spectrum of tensor perturbations has to be very blue and strongly violate the consistency relation, and therefore cannot be realized within the simplest single-field slow-roll models. However, scenarios leading to a blue spectrum have been studied in recent years. A by no means exhaustive list of examples includes the following: breaking of spatial~\citep{Endlich:2012pz,Cannone:2014uqa} or spatial and temporal diffeomorphism invariance~\citep{Ricciardone:2016lym}; axion-gauge field couplings~\citep{Dimastrogiovanni:2016fuu}; sizeable coupling of the inflaton to spin-2 fields~\citep{Iacconi:2019vgc}; higher curvature corrections to the effective gravitational action~\citep{Giare:2020plo}; modifications to gravity~\citep{Kobayashi:2010cm}; inflation in an Universe filled with an elastic medium~\citep{Gruzinov:2004ty}; an inflationary sound speed resonance~\citep{Cai:2020ovp}; or a string gas cosmology phase~\citep{Brandenberger:2006xi}.

\begin{figure}
\includegraphics[width=0.9\linewidth]{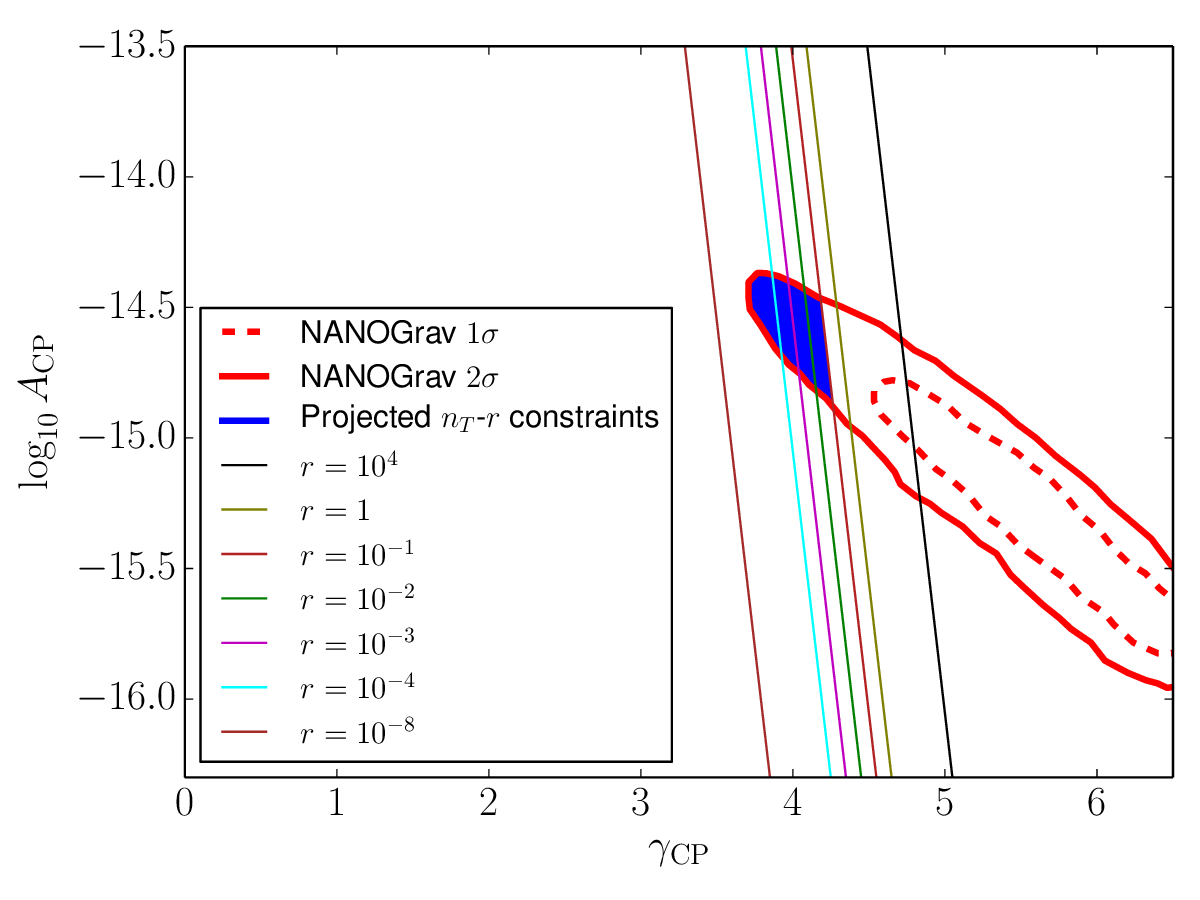}
\caption{Blue contour: $n_T$-$r$ contour of Fig.~\ref{fig:nanograv_fig1} projected onto the $\gamma_{_{\rm CP}}$-$A_{_{\rm CP}}$ parameter space. Dashed and solid red curves: $1\sigma$ and $2\sigma$ $\gamma_{_{\rm CP}}$-$A_{_{\rm CP}}$ contours obtained by NANOGrav. Solid diagonal lines:predictions for $\gamma_{_{\rm CP}}$ and $A_{_{\rm CP}}$ obtained for different fixed values of $r$ (see legend for color coding), while $n_T$ is varied.}
\label{fig:nanograv_fig2}
\end{figure}

It is instructive to understand where the previous constraints are coming from. One might be surprised by the fact that NANOGrav sets a \textit{lower} limit on $r$, in addition to the \textit{upper} limit from \textit{BKP}. In Fig.~\ref{fig:nanograv_fig2}, I project the $n_T$-$r$ constraints from Fig.~\ref{fig:nanograv_fig1} onto the $\gamma_{_{\rm CP}}$-$A_{_{\rm CP}}$ plane, with the $1\sigma$ and $2\sigma$ NANOGrav contours overlaid, and also plot the values of $\gamma_{_{\rm CP}}$ and $A_{_{\rm CP}}$ obtained for various fixed values of $r$, while $n_T$ is varied. From Fig.~\ref{fig:nanograv_fig2} one notices first of all that an inflationary SGWB is only able to account for a limited region of the NANOGrav parameter space. One notes that the contours at fixed $r$ are nearly vertical, and hence increasing (decreasing) $r$ approximately moves one to the right (left) in the $\gamma_{_{\rm CP}}$-$A_{_{\rm CP}}$ plane: therefore, the lower limit of $r \gtrsim {\cal O}(10^{-6})$ is driven by the leftmost edge in the NANOGrav $2\sigma$ contours. For any given value of $r$, there is a limited range of $n_T$ values for which the corresponding values of $\gamma_{_{\rm CP}}$ and $A_{_{\rm CP}}$ are consistent with the NANOGrav signal. This is reflected in the limited horizontal width of the blue-shaded region in Fig.~\ref{fig:nanograv_fig1}.

Taken at face value, the region of parameter space indicated by Fig.~\ref{fig:nanograv_fig1} appears to be consistent with current constraints on $n_T$ and $r$ from CMB measurements: see e.g.~\cite{Cabass:2015jwe,Giare:2019snj} for examples (although~\cite{Giare:2020vss} suggest that constraints on $n_T$ might be substantially tighter). However, GWs contribute to the energy density in the early Universe, behaving as extra radiation and contributing to the effective number of relativistic species $N_{\rm eff}$, which affects Big Bang Nucleosynthesis (BBN) and the CMB~\citep[see e.g.][]{Allen:1997ad,Smith:2006nka,Boyle:2007zx,Kuroyanagi:2014nba,Ben-Dayan:2019gll}. The SGWB contribution to $N_{\rm eff}$, $\Delta N_{\rm eff,SGWB}$, is given by:
\begin{eqnarray}
\int_{f_{\min}}^{f_{\max}} df\,\frac{\Omega_{\rm GW}(f)h^2}{f} \approx 5.6 \times 10^{-6} \Delta N_{\rm eff,SGWB}
\label{eq:deltaneff}
\end{eqnarray}
Extra contributions to $N_{\rm eff}$ are severely constrained by their impact on the damping of CMB higher acoustic peaks, and on light element abundances. Together, CMB and BBN roughly constrain $\Delta N_{\rm eff} \lesssim 0.4$~\citep[see e.g.][]{Aver:2015iza,Cooke:2017cwo,Aghanim:2018eyx,Vagnozzi:2019ezj,Hsyu:2020uqb,Aiola:2020azj,Mossa:2020gjc}.

Inserting the previous conservative constraint into Eq.~(\ref{eq:deltaneff}) determines BBN limits on $n_T$ and $r$. The lower integration limit is set $f_{\min}=10^{-10}\,{\rm Hz}$, approximately corresponding to the size of the comoving horizon at the time of BBN. On the other hand, the upper integration limit $f_{\max}$ is related to the maximum temperature reachable by the Universe, \textit{i.e.} the reheating temperature at the end of inflation $T_{\rm rh}$. For a given choice of $n_T$ and $r$, the BBN limit from Eq.~(\ref{eq:deltaneff}) is therefore a function of $f_{\max}$ and hence $T_{\rm rh}$.

In Fig.~\ref{fig:nanograv_fig1}, the dashed lines show the constraints arising from BBN limits on $N_{\rm eff}$ for different choices of $T_{\rm rh}$, with the region to the right of the lines excluded. For GUT-scale reheating ($T_{\rm rh} \approx 10^{15}\,{\rm GeV}$) and down to $T_{\rm rh} \approx 100\,{\rm TeV}$, BBN constraints exclude the entire region of parameter space required to explain NANOGrav. A large part of this parameter space becomes consistent with BBN when $T_{\rm rh}$ is as low as $1\,{\rm TeV}$, whereas full consistency is reached for reheating at the electroweak scale or lower ($T_{\rm rh} \lesssim 100\,{\rm GeV}$).~\footnote{See also the related work of~\cite{Kuroyanagi:2020sfw}, which appeared after the present manuscript was submitted to arXiv, and reached similar conclusions considering more specific scenarios.}

Models with very low reheating have been studied for instance in~\cite{Kawasaki:1999na,Kawasaki:2000en,Giudice:2000ex,Giudice:2000dp,Hannestad:2004px,Khoury:2011ii,Hasegawa:2019jsa,Hasegawa:2020ctq}. These scenarios are generally not easy to come about, and require either an unusually long period of slow-roll (thus a very flat potential), the existence of long-lived massive particles (such as a curvaton, the gravitino, or moduli), or scenarios alternative to inflation. Nonetheless, very low-reheating scenarios with $T_{\rm rh} \gtrsim {\cal O}({\rm MeV})$ can in principle be realized, and are thus worthy of consideration. In addition, $T_{\rm rh}$ could be as low as $\simeq 5\,{\rm MeV}$ without spoiling successful BBN~\citep{deSalas:2015glj,Gerbino:2016sgw}. In any case, $T_{\rm rh}$ is expected to be lowered even in the absence of new physics, by effects such as ``Higgs blocking''~\citep{Freese:2017ace,Litsa:2020rsm}.

An important caveat to my findings is $P_t(k)$ being treated as a pure power-law from CMB down to PTA scales. This approximation is widely used in the literature, but as we are approaching the point where it becomes possible to test inflation on such small scales, it is worth re-examining whether it is justified. Recently~\cite{Giare:2020vhn} have shown that higher-order terms in the power-law expansion of $P_t(k)$ can be extremely relevant when constraining inflation on small scales. Addressing this caveat requires estimating the theoretical uncertainties in the predictions for inflationary parameters in a way which is as potential-agnostic as possible, \textit{e.g.} through Monte Carlo inflation flow potential reconstruction~\citep[see e.g.][]{Kinney:2005in,Caligiuri:2014ola}. This very important question goes beyond the scope of my work, and I plan to return to it in the future.

\section{Conclusions}
\label{sec:conclusions}

The NANOGrav collaboration reported evidence for a stochastic common-spectrum process affecting pulsar timing residuals across its 12.5-year dataset. In this \textit{Letter}, I have examined whether the NANOGrav result can be explained by an inflationary SGWB, and if so what region of inflationary parameter space can explain the observation.

My key results are given in Fig.~\ref{fig:nanograv_fig1}, and indicate that the required primordial tensor power spectrum would have to be extremely blue ($n_T>0$) to comply with constraints on the tensor-to-scalar ratio $r$. The required region of $n_T$-$r$ parameter space can be brought in agreement with BBN constraints on the radiation energy density in the early Universe if the reheating scale is sufficiently low, $T_{\rm rh} \lesssim 100\,{\rm GeV}-1\,{\rm TeV}$. Therefore, for sufficiently low reheating scale, an inflationary interpretation of the possible first ever detection of a SGWB by NANOGrav is not excluded.

A caveat to my results is the assumption, widespread in the literature, of the primordial tensor power spectrum being a pure power-law across various decades in frequency. If confirmed, the hint for the possible discovery of a SGWB from NANOGrav would be an extraordinary milestone in GW astronomy. For this to occur, a convincing detection of spatial quadrupolar correlations is required.

\section*{Acknowledgements}

I thank William Giar\`{e}, Will Kinney, and Kai Schmitz for helpful discussions. I am supported by the Isaac Newton Trust and the Kavli Foundation through a Newton-Kavli fellowship, and acknowledge a College Research Associateship at Homerton College, University of Cambridge.

\section*{Data availability}

The data underlying this article will be shared upon request to the corresponding author.

\bibliographystyle{mnras}
\bibliography{NANOGrav}

\bsp
\label{lastpage}
\end{document}